\documentclass[useAMS,usenatbib]{mn2e}
\usepackage{times}

\newcommand{\jsix}{XTE J1650--500}
\newcommand{\xmm}{{\it XMM-Newton}}
\newcommand{\flux}{erg\,cm$^{-2}$\,s$^{-1}$}
\newcommand{\lum}{erg\,s$^{-1}$}

\title[\jsix\ in quiescence]{XMM-Newton observations of the black
hole X-ray transient XTE J1650--500 in quiescence}

\author[J. Homan et al.]
       {Jeroen Homan,$^1$\thanks{E-mail:jeroen@space.mit.edu} 
       Rudy Wijnands,$^2$ 
       Albert Kong,$^1$ 
       Jon M.\ Miller,$^{3}$              
       \newauthor          
       Sabrina Rossi,$^4$ 
       Tomaso Belloni,$^4$ 
       Walter H.G.\ Lewin\,$^1$ \\
$^1$MIT Kavli Institute for Astrophysics and Space Research, 70
Vassar Street, Cambridge, MA 02139, USA\\
$^2$Astronomical Institute "Anton Pannekoek", University of
Amsterdam, Kruislaan 403, 1098 SJ, Amsterdam, The Netherlands\\
$^3$The University of Michigan, 500 Church Street, Dennison 814, Ann Arbor, MI 48109-1042\\
$^4$INAF Osservatorio Astronomico di Brera, via E.\ Bianchi 46, 23807
Merate (LC), Italy}

\pagerange{\pageref{firstpage}--\pageref{lastpage}}

\begin{document}

\maketitle

\label{firstpage}

\begin{abstract} We report the result of an {\it XMM-Newton}
observation of the black-hole X-ray transient \jsix\  in quiescence.
The source was not detected and we set upper limits on the 0.5--10
keV luminosity of $0.9-1.0\times10^{31}$ \lum\ (for a newly derived
distance of 2.6 kpc). These limits are in line with the quiescent
luminosities of black-hole X-ray binaries with similar orbital
periods ($\sim$7--8 hr).

\end{abstract}


\section{Introduction}

\jsix\ is an X-ray transient that was discovered with the {\it Rossi
X-ray Timing Explorer (RXTE)} in 2001 \citep{re2001}, during the only
outburst of the source detected thus far. Based on its spectral and
variability behavior \jsix\ can be classified as a black hole
candidate X-ray binary \citep[see
e.g.][]{katoro2003,hoklro2003,rohomi2004}. Optical observations of
the source revealed an orbital period of 0.3205 days and a mass
function $f(M)=2.73\pm0.56\,M_\odot$  \citep{ormcre2004}. Combined
with a lower limit of $50^\circ\pm3^\circ$ on the inclination, this
translates into an upper limit on the black hole mass of
$7.3\,M_\odot$, with a most likely mass of $\sim$4 $M_\odot$ 
\citep{ormcre2004}. Observations with \xmm\ during outburst revealed
a red-shifted Fe K-alpha line \citep{mifawi2002}, which suggests that
the compact object is a rapidly spinning black hole. \jsix\ has been observed
at luminosities (0.5--10 keV) ranging from $\sim4\times10^{33}$
\lum\  to $\sim8\times10^{36}$ \lum\ (for an assumed distance of 2.6
kpc, see Appendix \ref{app:dist}). At low luminosities, when it was in the
hard state, the source showed two  properties that have not been
shown by any other transient black hole X-ray binary. Oscillations
with a period of 14 days were observed, which had similarities to the
longterm oscillations observed in the transient millisecond-second
X-ray pulsar SAX J1808.4-3658 \citep{wi2004}. Also, six X-ray flares
were observed \citep{tokaco2003}. Using {\it Chandra} and {\it RXTE}
data, \citet{tokaka2004} observed spectral softening towards low
X-ray luminosities, with the spectral power-law index evolving from
1.66$\pm$0.05 at $9\times10^{34}$ \lum\ (1--9 keV) to 1.91$\pm$0.13
at $1.5\times10^{34}$ \lum. Prior to our work, the source was not
observed in quiescence.

In this paper we report the results of an \xmm\ observation of \jsix,
performed when the source was believed to be in quiescence, more than
two and a half years after the source was last observed with {\it
RXTE} and {\it Chandra} in 2002. About fourteen black-hole X-ray
binaries (BHXBs) have so far been observed in quiescence  (most of
them with {\it Chandra} and {\it XMM-Newton}), with detected 0.5--10
keV luminosities ranging from $4\times10^{30}$ \lum\ to
$1\times10^{33}$ \lum\ \citep[see \citet{tocofe2003} for a complete
list of observed sources]{habala2003}. These studies of quiescent
BHXBs are playing a central role in the debate on the existence of
black-hole event horizons. As noted by various authors
\citep{nagamc1997,meesna1999,gamcna2001}, BHXBs have substantially
lower quiescent X-ray luminosities than neutron star X-ray binaries
(NSXBs) with similar orbital periods. Usually, the neutron-star
systems have a relatively bright soft, thermal component which is
lacking in the BHXBs (but sometimes also in NSXBs
[\citet{castga2002,wihepo2005}]). This soft component is interpreted
as coming from the neutron-star surface, either due to the cooling of
the neutron star and/or due to a crustal heating by residual
accretion onto the neutron star's surface. The absence of such a
thermal component and the lower quiescent X-ray luminosities are
therefore often interpreted as an indication for the presence of an
event horizon in BHXBs. However, since the source of quiescent X-ray
emission in both the BHXBs and the NSXBs is still under debate, this
conclusion remains uncertain.

Several sources of quiescent X-ray emission have been proposed for
BHXBs. It has been suggested \citep[see e.g.][and references
therein]{nagamc1997,nabamc1997,habala2003} that advection dominated
accretion flows (ADAFs) can explain the optical/X-ray ratios and
X-ray spectra seen in quiescence. These flows also provide a possible
explanation for the luminosity difference between black-hole and
neutron-star X-ray transients in quiescence. \citet{biru2000}
suggested that coronal emission from rapidly rotating companion stars
might be responsible for the quiescent X-ray luminosity. However, it
was shown by \citet{komcga2002} that the X-ray spectra of five of the
six quiescent BHXBs they studied were inconsistent with those of
stellar coronae. Finally, assuming that the hard state of BHXBs
extends all the way to quiescence, another possibility is X-ray
emission from a jet \citep{mafafe2001,manoco2003}. Regardless of the
exact mechanism for the X-ray emission, \cite{fegajo2003} suggested
that at very low mass accretion rates BHXBs should enter a
`jet-dominated' state in which most of the energy is released in a
jet outflow instead of X-rays from the accretion flow. At similarly
low mass accretion rates, the energy release from NSXBs is not
expected to be  dominated by jets. 

In addition to the luminosity difference between quiescent BHXBs and
NSXBs,  there also appears to be a positive correlation between the
quiescent luminosities and orbital periods of BHXBs \citep[see
e.g.][]{habala2003}. Based on a comparison with sources that have
similar orbital periods one would expect a quiescent X-ray luminosity
for \jsix\ of $\sim10^{30}$--$10^{31}$ \lum.

\section{analysis and results}

\jsix\ was observed with \xmm\ from March 6 2005 14:59 UT until 03:42
UT the following day (Obs-Id: 0206640101). For this paper we analyzed
pipeline-production data from the EPIC-pn and two EPIC-MOS
instruments, using SAS version 6.1.0. All three instruments were used
with a medium thickness filter. For each of the three instruments
background light curves were produced from all CCDs combined,
selecting only photons above 10 keV (with PATTERN=0 and FLAG=0). The
resulting light curves showed very strong flaring for $\sim$65\%
(MOS) and $\sim$75\% (pn) of the observation. Intervals free of
flaring were singled out by selecting 50 s time bins with count rates
below 0.65 count\,s$^{-1}$ (pn) or below 0.2 counts\,s$^{-1}$ (MOS).
The resulting good time intervals were applied to the CCD on which
the source was located, using the standard selection criteria
(PATTERN$\leq$4 for EPIC-pn, PATTERN$\leq$12 for MOS, and FLAG=0).
This resulted in effective exposure times of 12 ks (pn) and 18 ks
(MOS).

Images were produced in the 0.5--10 keV band with a binning that
resulted in pixels of 5\arcsec, close to the full-width-at-half-maximum of
the point spread function (PSF) of the EPIC cameras. None of the images
showed an apparent excess of photons at the location of the source.
Images in narrower energy bands across the 0.5--10 keV range did not
show an obvious source detection either.

Next, we ran the SAS task {\tt edetect\_chain} (with default input
parameters) simultaneously on the three 0.5--10 keV images. No source
was detected at the position  of \jsix\ (R.A.=16:50:01.0,
Dec=-49:57:45). For the faintest source\footnote{XMMU
J165002.6-495333, a total of 49$\pm$11 source counts were detected
from this source (pn and MOS combined)} detected with  {\tt
edetect\_chain} we calculated the flux using PIMMS,
assuming\footnote{The value of $N_H$ was obtained from spectral fits
to \xmm\ data of \jsix\ in outburst \citep{mifawi2002} and is
consistent with values obtained from radio measurements
\citep{dilo1990}.} an $N_H$ of $5.3\times10^{21}$ atoms\,cm$^{-2}$
and a power-law spectrum with indices ($\Gamma$) varying between 1.5
and 2.5.  For EPIC-MOS we found unabsorbed fluxes  between
2.0(7)$\times10^{-14}$ \flux\ ($\Gamma=2.5$) and
2.2(8)$\times10^{-14}$ \flux\ ($\Gamma=1.5$) and for EPIC-pn we found
fluxes between 3.1(9)$\times10^{-14}$ \flux\ ($\Gamma=2.5$) and
3.5(9)$\times10^{-14}$ \flux\ ($\Gamma=1.5$). These numbers serve as
conservative upper limits on the flux of \jsix.

 Alternative upper limits were obtained from the 0.5--10 keV images
by extracting the counts from a 20\arcsec radius circle centered on
the position of \jsix. This was done using the tool {\tt funcnts}
from the FUNTOOLS package, which also extracted counts from a large
pre-defined source-free background region and corrected for
differences in the exposure map values between the source and
background region. The numbers of counts (source + background) within
this region were 22, 18 and 35, for MOS1, MOS2, and pn, respectively.
Using the tables in \citet{ge1986} this leads to single-sided
3$\sigma$ upper limits of 40.1, 34.8, and 56.7 on the total counts.
The expected background counts for the three source regions were,
26.9, 21.5 and, 44.1, respectively, with errors of less than 5\%.
Correcting for background and using the average exposure map values
(17.7 ks, 15.6 ks, and 10.3 ks), this gives upper limits on the
source count rate of 7.5$\times10^{-4}$, 8.5$\times10^{-4}$ and
1.23$\times10^{-3}$ counts\,s$^{-1}$.  Using PIMMS, correcting for
encircled energy (from PSF integration\footnote{We extracted from a
20\arcsec radius, while PIMMS expects count rates from a 15\arcsec
region. Using figures from the XMM Users' Handbook, we estimate the
correction factors to be 0.92 (mos) and 0.91 (PN).}), and assuming an
$N_H$ of $5.3\times10^{21}$ atoms\,cm$^{-2}$, we get the following
(unabsorbed) flux upper limits for power-law spectra:
$1.8-2.0\times10^{-14}$ \flux\ (MOS1), $2.1-2.3\times10^{-14}$ \flux\
(MOS2), and $1.1-1.3\times10^{-14}$ \flux\ (pn), with the high values
corresponding to $\Gamma=1.5$ and the low values to $\Gamma=2.5$.

\section{Conclusions}

We have observed the black hole X-ray transient \jsix\ in quiescence.
The source was not detected. Assuming a distance of 2.6 kpc (see
Appendix \ref{app:dist}), the most stringent upper limits on 0.5--10
keV luminosity are $0.9-1.0\times10^{31}$ \lum\ (depending on the
assumed shape of the spectrum), which corresponds to
$\sim1.8\times10^{-8}$ times the Eddington luminosity ( $L_{Edd}$)
for a black-hole mass of 4$M_\odot$ \citep{ormcre2004}. These values
are similar to the quiescent luminosities of other BHXBs with similar
orbital periods, even for a distance as low as 2 kpc (which is
consistent with our estimate). Our luminosity upper limits follow the
observed relation between orbital period and quiescent luminosity,
although, as mentioned also by others, more systems with long periods
need to be observed to firmly establish the existence of such a
relation. 

Longer observations are needed to detect \jsix\ in quiescence or
provide more stringent and useful upper limits on its luminosity,
preferably with {\it Chandra}, which suffers relatively less from
background flaring than \xmm\ and has a much lower non-flaring
background than \xmm .

\section*{Acknowledgments}

The authors would like to thank the referee for carefully reading the
manuscript and for spotting an error in our flux calcualtions. The
results presented in this paper are based on observations obtained
with XMM-Newton, an ESA science mission with instruments and
contributions directly funded by ESA Member States and NASA. JH and
WHGL acknowledge generous support from NASA.

\appendix

\section{Distance estimate}\label{app:dist}

Here we derive a distance estimate for \jsix, based on the luminosity
at which the source made the transition from the spectrally soft
state back to the spectrally hard state at the end of its 2001
outburst. \citet{ma2003} studied such transitions in six black hole
X-ray transients and found that they occur in a narrow rang of
luminosity: 1--4\% of the Eddington luminosity ($L_{Edd}$), with an
average of 2\% of $L_{Edd}$. The definition of the 'transition to the
hard state' is not entirely clear from \citet{ma2003}, but we pick an
{\it RXTE} observation of \jsix\ in which the spectral power-law
index was around 2, similar to the {\it RXTE} observation that
\citet{ma2003} used for XTE J1550--564. Note that  \citet{ma2003}
assumed a spectral index of 1.8 and an exponential cutoff at 200 keV
for the bolometric correction (to the 0.5 keV to 10 MeV range). No
observation with such a spectral index was present in the coverage of
the soft-to-hard state transition in \jsix, with a spectral index of
2 being the closest.  We analyzed {\it RXTE} observation
60113-01-39-02 and fitted the 3--100 keV spectrum with a simple
phenomenological model consisting of a disk blackbody, a cut-off
power law, and a smeared edge, with $N_H$ fixed to $5.3\times10^{21}$
atoms\,cm$^{-2}$. The unabsorbed flux, extrapolated to the 0.5 keV to
10 MeV range, is $1.4\times10^{-8}$ \flux. Assuming that this flux
corresponds to 2\% of $L_{Edd}$, which is  $\sim1.1\times10^{37}$
\lum\ for a 4 $M_\odot$ black hole, we derive a distance of
2.6$\pm$0.7 kpc. This is consistent with the distance of 3 kpc
derived by \citet{cofeto2004}, following a similar method. The main
contributors to errors on the distance are the spread in the fraction
of $L_{Edd}$ at which the soft-to-hard transition occurs (fractional
error $\sim$50\%) and uncertainties introduced by extrapolating well
outside the spectral fit range ($\sim$30\%).

%\bibliographystyle{mn2e}
%\bibliography{1650_xmm}

\label{lastpage}

\end{document}